\documentclass[conference]{IEEEtran}
\IEEEoverridecommandlockouts
\usepackage{amsmath}
\usepackage{amsfonts}
\usepackage{bbding}
\usepackage{amssymb}
\usepackage{array}
\usepackage{subfigure}

\usepackage{graphicx}
\usepackage{subfigure}
\usepackage[named]{algo}
\usepackage{algorithmic}
\usepackage{psfrag}
\usepackage{xfrac}
\usepackage{stfloats}
\usepackage[compress]{cite}
\makeatletter
\renewcommand{\citepunct}{,\penalty\@m\hskip.13emplus.1emminus.1em}
\renewcommand{\citedash}{\hbox{--}\penalty\@m}
\makeatother
\usepackage{setspace}
\usepackage{color}
\allowdisplaybreaks

\usepackage{amsthm}
\usepackage{stfloats}

\newtheorem{pro}{Property}

\usepackage{hyperref}
\usepackage{bm}
\begin{document}
\title{Energy-Efficient Resource Allocation for Ultra-reliable and Low-latency Communications}

\author{
\IEEEauthorblockN{{Chengjian Sun, Changyang She and Chenyang Yang}} \vspace{0.0cm}
\IEEEauthorblockA{School of Electronics and Information Engineering,\\ Beihang University, Beijing, China\\
Email: \{sunchengjian,cyshe,cyyang\}@buaa.edu.cn}\vspace{-0.6cm}
}
\maketitle
\begin{abstract}
Ultra-reliable and low-latency communications (URLLC) is expected to be supported without compromising the resource usage efficiency. In this paper, we study how to maximize energy efficiency (EE) for URLLC under the stringent quality of service (QoS) requirement imposed on the end-to-end (E2E) delay and overall packet loss, where the E2E delay includes queueing delay and transmission delay, and the overall packet loss consists of queueing delay violation, transmission error with finite blocklength channel codes, and proactive packet dropping in deep fading. Transmit power, bandwidth and number of active antennas are jointly optimized to maximize the system EE under the QoS constraints. Since the achievable rate with finite blocklength channel codes is not convex in radio resources, it is challenging to optimize resource allocation. By analyzing the properties of the optimization problem, the global optimal solution is obtained. Simulation and numerical results validate the analysis and show that the proposed policy can improve EE significantly compared with existing policy.
\end{abstract}

\section{Introduction}
Ultra-reliable and low-latency communications (URLLC) is crucial to enable mission critical applications such as autonomous vehicle communications, factory automation and haptic communications \cite{Factory2015Yilmaz,Meryem2016Tactile}.

To ensure the low end-to-end (E2E) delay including uplink (UL) and downlink (DL) transmission delay, coding and processing delay, queueing delay, and routing delay in backhaul and core networks, short frame structure becomes necessary \cite{Petteri2015A}, queueing delay should be controlled \cite{Adnan2016Towards}, and network architecture needs to be updated. To ensure the high reliability characterized by the overall packet loss probability, including transmission errors in UL and DL transmissions, packet loss due to queueing delay violation and proactive packet dropping in deep fading \cite{She2016CrossLayer}, together with the short delay guarantee, channel coding with finite blocklength ought to be used \cite{Yury2014Quasi}, and various diversity techniques are critical \cite{Beatriz2015Reliable}. While supporting the stringent quality of service (QoS) of URLLC is not an easy task for radio access networks, the resource usage efficiency should not be compromised.

Energy efficiency (EE) is a key performance metric of the fifth generation (5G) mobile communications, which has been extensively studied (please see \cite{YangGR2015,Wenjuan2016TWC} and references therein). However, the methods in literatures can hardly be extended to URLLC mainly due to the following reasons. First, both the EE and QoS depend on the achievable rate. The Shannon's capacity used in existing studies cannot characterize the maximal achievable rate with given transmission error, and hence can no longer be applied to URLLC. Second, owing to the very short delay bound, existing tools of analyzing queueing delay may not be applicable to URLLC.

An early attempt to design energy efficient URLLC in \cite{She2016EEtactile} showed that effective bandwidth \cite{EB} can be used to characterize queueing delay for Poisson process. More recently, the study in \cite{She2016CrossLayer} validated that effective bandwidth can be applied for arrival processes that are more bursty than Poisson process. By further introducing a simple approximation to the achievable rate with finite blocklength channel codes, the transmit power and bandwidth allocation policy was optimized with closed-form solution to maximize system EE under constraints on transmission and queueing delays and corresponding packet loss components \cite{She2016EEtactile}. However, such an approximation is not always accurate, since the signal-to-noise ratio (SNR) and the blocklength of different users change in a wide range.
In practice, circuit power consumption highly depends on the number of active antennas, which was not optimized in \cite{She2016EEtactile}. How to jointly allocate transmit power and configure bandwidth and antenna to maximize the EE for URLLC with accurate achievable rate approximation deserves further study.

In this paper, we optimize EE of URLLC. Since around $80$\% energy consumption in cellular networks comes from BSs \cite{YangGR2015}, we only consider the energy consumed at the BSs. We jointly optimize transmit power, bandwidth and the number of active antennas to maximize EE under the QoS constraints. To capture the basic idea, we focus on DL resource allocation. Different from \cite{She2016EEtactile}, an accurate approximation of the achievable rate with finite blocklength channel codes in \cite{Yury2014Quasi} will be applied, which however leads to very complicated QoS constraints. By further introducing an approximation and an upper bound, the QoS constraints are expressed in closed-form, but are still non-convex in both transmit power and bandwidth. By analyzing the properties of the optimization problem, the global optimal resource allocation policy is obtained. Simulation and numerical results show that the proposed policy can satisfy the QoS requirement and provide remarkable EE gain over existing policy.


\section{System Model}

\subsection{System and Traffic Models}
\begin{figure}[htbp]
	\vspace{-0.2cm}
	\centering
	\begin{minipage}[t]{0.48\textwidth}
	\includegraphics[width=1\textwidth]{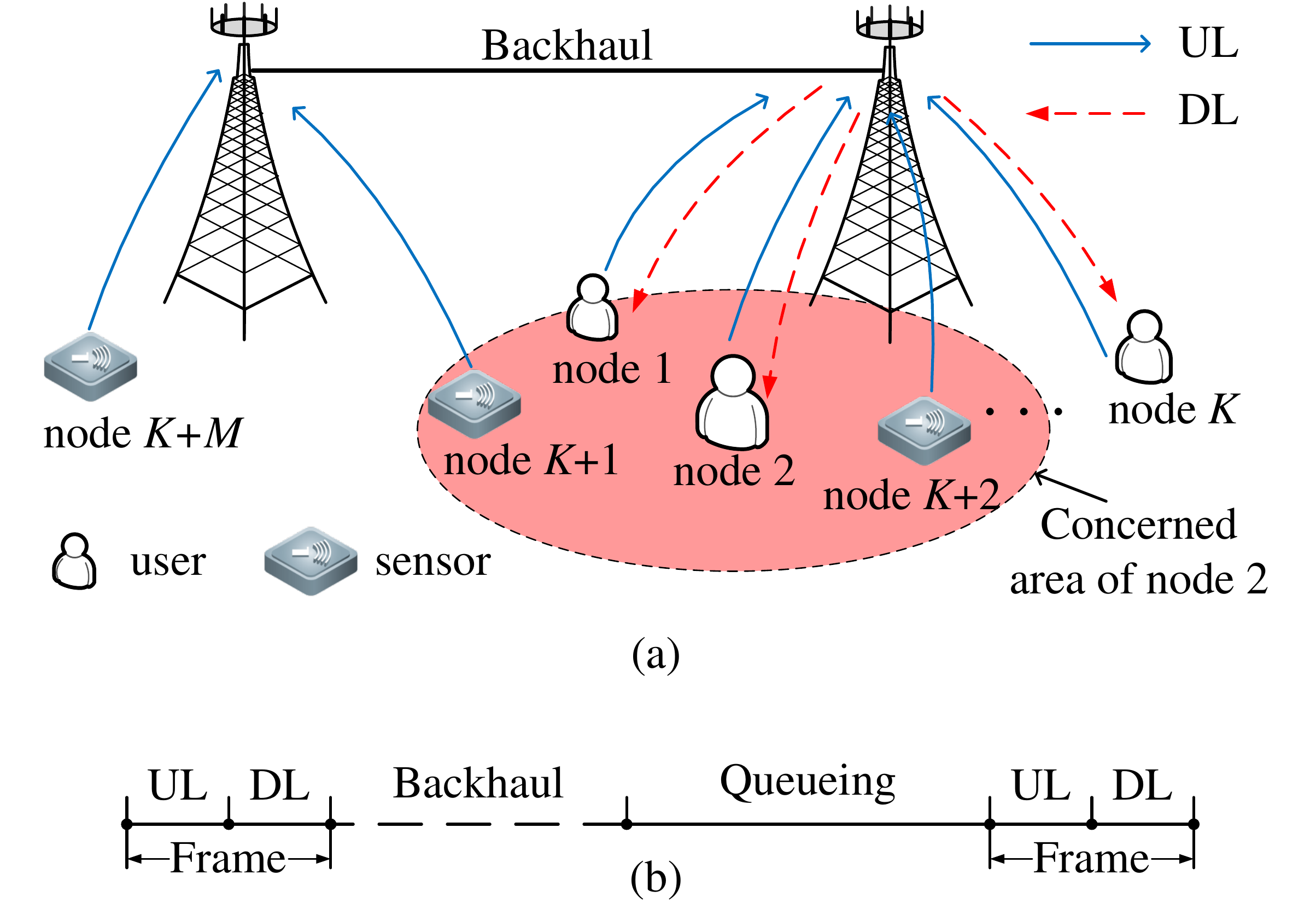}
	\end{minipage}
	\vspace{-0.3cm}
	\caption{System model and E2E delay.}
	\label{fig:smodel}
	\vspace{-0.4cm}
\end{figure}

Consider a cellular system, where each BS with $N_\mathrm{t}$ antennas serves $K\!+\!M$ single-antenna nodes. The nodes are divided into two types. The first type of nodes are $K$ users, which upload and download packets to and from the BS. The second type of nodes are $M$ sensors, which only upload packets. We consider the local communication scenario where each user only needs the packets from the nodes located in a few adjacent cells. All the nodes upload their own messages in short packets to their accessed BSs in UL. If a BS receives the messages from the nodes (e.g., node $K+1$ in Fig. \ref{fig:smodel}(a)) that are required by the users accessed to an adjacent BS (e.g., node 2), the BS forwards the required messages to the adjacent BS via backhaul, and then the adjacent BS transmits the relevant messages to the target users. Such a scenario can be found in autonomous vehicle communications, smart factory and some augmented reality applications \cite{Meryem2016Tactile}, where the propagation delay is negligible, and the fiber backhaul latency is very short (around 0.1 ms \cite{Tony2015Delay}). For long distance communication scenarios, the E2E delay ranges from $1$~ms to $100$~ms, depending on the use cases \cite{Philipp2017Latency}. Nevertheless, according to \cite{3GPP2016Scenarios}, the latency in radio access network should not exceed $1$~ms. We reserve the delay for backhaul transmission for simplicity, and focus on the latency in radio access networks.

Time is discretized into frames. Each frame has duration $T_\mathrm{f}$, which includes UL and DL transmissions.
We consider frequency reuse and frequency division multiple access to avoid interference.

\vspace{-0.4cm}

\subsection{Channel Model}
We consider typical use cases in URLLC, where the E2E delay requirement $D_\mathrm{max}$ (and hence channel coding blocklength and  the queueing delay) is shorter than the channel coherence time.
Since the data rate for URLLC is not high, it is reasonable to assume that the bandwidth for each user is less than the channel coherence bandwidth, and then the channel is flat fading. Denote the average channel gain and the normalized channel coefficient for the $k$th user as $\alpha_k$ and $\bm{h}_k \!\in\! {\mathbb{C}^{N_\mathrm{t}\!\times\!1}}$, respectively, where the elements of $\bm{h}_k$ are independent and identically complex Gaussian distributed with zero mean and unit variance.

When both $\alpha_k$ and $\bm{h}_k$ are known at the BS, for a given transmission error probability $\varepsilon_k^c$, the achievable packet rate with finite blocklength can be accurately approximated by \cite{Yury2014Quasi}
\begin{align}\label{eq:sk}
	s_k \approx \frac{ \phi W_k}{u \ln{2}} \left\{\ln\left[1+\frac{\alpha_k P_k^\mathrm{t} g_k}{ N_0 W_k}\right] -
				\sqrt{\frac{V_k}{\phi W_k}}Q_\mathrm{G}^{-1}\left({\varepsilon_k^c}\right)\right\},
\end{align}
where $P_k^\mathrm{t}$ and $W_k$ is the transmit power and bandwidth allocated to the $k$th user, respectively, $u$ is the number of bits contained in each packet, $\phi \!\in\! (0,T_\mathrm{f})$ is the time that can be used for DL transmission in one frame, $g_k \!=\! \bm{h}_k^\mathrm{H} \bm{h}_k$, $[\cdot]^\mathrm{H}$ denotes conjugate transpose, $N_0$ is the single-side noise spectral density, $Q_\mathrm{G}^{-1} (x)$ is the inverse of the Gaussian Q-function, and $V_k$ is the channel dispersion given in \cite{Yury2014Quasi}
\begin{align}	\label{eq:dispersion}
	V_k=1-\frac{1}{\left[1+\frac{\alpha_k P_k^\mathrm{t} g_k}{N_0 W_k }\right]^2}.
\end{align}

\subsection{QoS Requirement}
As shown in Fig. \ref{fig:smodel}(b), the E2E delay $D_\mathrm{max}$ consists of UL transmission delay, backhaul latency, queueing delay and DL transmission delay. Since the packet size is very small (e.g., $20$~bytes \cite{3GPP2016Scenarios}), we assume that both UL and DL transmission for a packet can be finished with given error probability in one frame without retransmission, i.e., the transmission delay can be ensured by the properly designed frame duration. For simplicity, assume that the backhaul delay is within one frame. Then, to ensure the E2E delay, the queueing delay should be $D_\mathrm{max}^q \!=\! D_\mathrm{max} \!-\! 2 T_\mathrm{f}$.

If the queueing delay of a packet is longer than the delay bound $D_\mathrm{max}^q$, then it has to be dropped reactively. Denote the queueing delay violation probability as $\varepsilon_k^q$. Since the queueing delay is shorter than channel coherence time in typical scenarios, the optimal transmit power allocated to ensure $(D_\mathrm{max}^q,\varepsilon_k^q)$ is channel inverse and could become unbounded  \cite{She2016CrossLayer}.
To guarantee the overall packet loss with finite transmit power, we can discard some packets proactively\cite{She2016CrossLayer}, i.e., drop some packets during deep channel fading even when their queueing delay is less than $D_\mathrm{max}^q$. Denote proactive packet dropping probability as $\varepsilon_k^h$. Then, the three packet loss components should satisfy $\varepsilon_k^c \!+\! \varepsilon_k^q \!+\! \varepsilon_k^h \!\leq\! \varepsilon_\mathrm{D}$ to ensure the overall reliability, where $\varepsilon_k^c$ is the DL transmission error probability and the UL transmission error probability has been subtracted from $\varepsilon_\mathrm{D}$.

\subsection{Energy Efficiency and Power Model}
EE is defined as the ratio of the average data rate to the average total power consumed at the BSs. By setting the frequency reuse factor as $1/3$, there is no strong interference and resource allocation policy of one BS does not depend on that of the other BSs. Therefore, maximizing the network EE is equivalent to maximizing the EE of each cell, which is
\begin{align}	\label{eq:EE}
	\eta=\frac{ \left(1-\varepsilon_\mathrm{D}\right) \sum_{k=1}^{K} \mathbb{E}\left\{\sum_{i\in{\mathbb{A}_k}}a_i(n)\right\} }
			  { \mathbb{E}\left\{P_\mathrm{tot}\right\} },
\end{align}
where $\mathbb{A}_k$ is the set of indices of nodes that lie in the concerned area of user $k$, $a_i(n)$ is the number of packets uploaded by node $i$ in the $n$th frame, and $\mathbb{E}\left\{P_\mathrm{tot}\right\}$ is the average total power consumed by the BS, which can be simplified as  \cite{Bjorn2015A},
\begin{align}	\label{eq:AvePtot}
	\mathbb{E}\left\{P_\mathrm{tot}\right\} = \frac{1}{\rho} \sum_{k=1}^K {\mathbb{E}\left\{P_k^\mathrm{t}\right\}} + P^\mathrm{ca} N_\mathrm{t} + P_0^\mathrm{c},
\end{align}
where $\rho\!\in\!{(0,1]}$ is the power amplifier efficiency, $P_k^\mathrm{t}$ is the transmit power allocated to user $k$, $P^\mathrm{ca}$ is the circuit power consumed by each antenna for transmission packets in DL and receiving packets from UL, and $P_0^\mathrm{c}$ is the fixed circuit power independent of the number of antennas.

\section{Resource Allocation Optimization}
In this section, we optimize the resource allocation that maximizes the EE meanwhile satisfies the QoS requirement. Since the nominator in \eqref{eq:EE} approximately does not depend on the resource allocation (because $1-\varepsilon_\mathrm{D} \approx 1$), maximizing EE is equivalent to minimizing the average total power consumption.

\subsection{Problem Formulation}
Denote the maximal transmit power of the BS as $P^\mathrm{t}_{\max}$. Then, the transmit power allocated to the users should satisfy $\sum_{k=1}^{K}{P^\mathrm{t}_k} \!\leq\! P^\mathrm{t}_{\max}$. With this constraint, the power allocated to each user depends on the channels of other users. As a result, it is hard to obtain the average transmit power of each user in closed-form. In order to find a closed-form expression of $\mathbb{E}\left\{P_\mathrm{tot}\right\}$ to facilitate optimization, we introduce a maximal transmit power constraint for each user as $P^\mathrm{t}_k \!\leq\! P^\mathrm{th}_k$, where $\sum_{k=1}^{K}{P^\mathrm{th}_k} \leq P^\mathrm{t}_{\max}$.

The minimal constant service rate required to ensure the queueing delay requirement $(D^q_{\max},\varepsilon_k^q)$ is the effective bandwidth of the arrival process \cite{EB}. For a Poisson process with arrival packet rate $\lambda_k$, the effective bandwidth is \cite{She2016CrossLayer},
\begin{align}\label{eq:EB}
    E_k^{B} = \frac{T_{\rm f}\ln(1/\varepsilon_k^q)}{D^q_{\max} \ln \left[1+\frac{T_\mathrm{f}\ln(1/\varepsilon_k^q)}{\lambda_k D^q_{\max}}\right]}.
\end{align}
Then, the queueing delay requirement can be satisfied if \cite{EB}
\begin{align}\label{eq:reqSk}
    s_k \geq E_k^{B}.
\end{align}
For other arrival processes, the expressions of $E_k^{B}$ will differ, but the proposed method to optimize resource allocation is still applicable.

By substituting the achievable packet rate (i.e., service rate) in \eqref{eq:sk} and effective bandwidth in \eqref{eq:EB} into \eqref{eq:reqSk}, the required SNR to guarantee both $\varepsilon_k^c$ and $(D^q_{\max},\varepsilon_k^q)$ should satisfy
\begin{align}	\label{eq:SNR}
	\gamma_k \geq \exp\left\{\frac{l_k(\varepsilon_k^q)}{W_k}+\frac{v_k(\varepsilon_k^c)}{\sqrt{W_k/V_k}}\right\}-1,
\end{align}
where $l_k(\varepsilon_k^q)\!=\!\frac { T_\mathrm{f} u \ln{2} \ln{\left(1/\varepsilon_k^q\right)} }{ \phi D_\mathrm{max}^q \ln{\left[1+\frac{T_\mathrm{f} \ln{\left({1}/{\varepsilon_k^q}\right)}}{D_\mathrm{max}^q \lambda_k}\right]} }$, and $v_k(\varepsilon_k^c)\!=\!Q_\mathrm{G}^{-1} \left(\varepsilon_k^c\right)/{\sqrt{\phi}}$.

According to \eqref{eq:dispersion}, we have $V_k \leq 1$. Since the right hand side of \eqref{eq:SNR} increases with $V_k$, the QoS requirement can still be satisfied with  $V_k\!=\!1$. By substituting $V_k\!=\!1$ into \eqref{eq:SNR}, the right hand side does not depend on $\gamma_k$, and then we obtain a conservative QoS constraint in closed-form.

To satisfy \eqref{eq:SNR} with $V_k\!=\!1$ and the maximal transmit power constraint, $P_k^\mathrm{t}$, depends on $g_k$ according to
\begin{align}	\label{eq:Pt1}
	P_k^\mathrm{t}=\begin{cases}
		P_k^\mathrm{th}, &\quad \text{if } g_k < g_k^\mathrm{th},\\
		\frac{N_0 W_k \gamma_k}{\alpha_k g_k}, &\quad \text{if } g_k \geq g_k^\mathrm{th},
	\end{cases}
\end{align}
where $g_k^\mathrm{th} \!\triangleq\! \frac {N_0 W_k \gamma_k} {\alpha_k P_k^\mathrm{th}}$ is a threshold.

When the channel gain is lower than $g_k^\mathrm{th}$, the required SNR to ensure QoS cannot be achieved. In this scenario, we can proactively discard several packets without transmission, and then control the overall packet loss/dropping probability. The achievable packet rate with transmit power $P_k^\mathrm{th}$ can be obtained by substituting $P_k^\mathrm{t} \!=\! P_k^\mathrm{th}$ into \eqref{eq:sk}, and is denoted as $s_k^\mathrm{th}$. Then, the number of packets that are discarded when $g_k < g_k^\mathrm{th}$ is \cite{She2016CrossLayer},
\begin{align}   \label{dk}
    d_k(n) = \min(T_{\rm f}E_k^B-s_k^\mathrm{th},Q(n)),\; \text{if} \; Q(n) > 0,
\end{align}
where $Q(n)$ is the number of packets waiting in the queue at the beginning of the $n$th frame.

From the derivation in \cite{She2016CrossLayer}, the packet dropping probability can be approximated as
\begin{align}   \label{eq:UB}
    \varepsilon_k^h &\triangleq \frac{{\mathbb{E}}[d_k(n)]}{{\lambda_k}} \approx B_{N_\mathrm{t}}\left( g_k^\mathrm{th} \right),
\end{align}
where
\begin{align}   \label{eq:B}
    B_{N_\mathrm{t}}\left( g_k^\mathrm{th} \right)\triangleq\int_0^{g_k^\mathrm{th}}{\left[1-\frac {\ln{\left(1+\frac{g \gamma_k}{g_k^\mathrm{th}}\right)}} {\ln{\left(1+\gamma_k\right)}}\right] f_{N_\mathrm{t}}\left(g\right) \mathrm{d}g}.
\end{align}

According to \eqref{eq:Pt1}, the average transmit power for the $k$th user can be expressed as
\begin{align}	\label{eq:AvePt1}
\begin{split}
	\mathbb{E}\!\left\{P_k^\mathrm{t}\right\}
		\!=\!\int_0^{g_k^\mathrm{th}}\!\! {P_k^\mathrm{th} f_{N_\mathrm{t}}\!\left(g\right) \mathrm{d}g} \!+\!
		   \int_{g_k^\mathrm{th}}^{\infty}\!\! {\frac{N_0 W_k \gamma_k}{\alpha_k g} f_{N_\mathrm{t}}\!\left(g\right) \mathrm{d}g},
\end{split}
\end{align}
where $f_{N_\mathrm{t}}\!\left( g \right) \!=\! \frac{g^{{N_\mathrm{t}}-1}}{\left( {N_\mathrm{t}}-1 \right)!} e^{-g}$ is the probability density function of $g$. By substituting $f_{N_\mathrm{t}}\!\left( g \right)$ into \eqref{eq:AvePt1}, we obtain
\begin{align}   \label{eq:AvePt2}
	\mathbb{E}\left\{P_k^\mathrm{t}\right\} = \frac{N_0 W_k \gamma_k}{\alpha_k \left({N_\mathrm{t}}-1\right)}
											\left[ 1-F_{N_\mathrm{t}}\left(g_k^\mathrm{th}\right) \right],
\end{align}
where
\begin{align}
	F_{N_\mathrm{t}}\!\left(g_k^\mathrm{th}\right)
        \!=\!& \left(N_\mathrm{t}\!-\!1\right)\!\int_0^{g_k^\mathrm{th}}\!\!{\left(\!\frac{1}{g}\!-\!\frac{1}{g_k^\mathrm{th}}\!\right) \! f_{N_\mathrm{t}}\!\left(g\right) \mathrm{d}g} \nonumber \\
        \!=\!& \int_0^{g_k^\mathrm{th}}\!\!{\left(\!1\!-\!\frac{g}{g_k^\mathrm{th}}\!\right) \!f_{N_\mathrm{t}\!-\!1}\!\left(g\right) \mathrm{d}g} \label{eq:F} \\
        \!=\!& \left(\!1\!-\!\frac{{N_\mathrm{t}}\!-\!1}{g_k^\mathrm{th}}\!\right) e^{-g_k^\mathrm{th}} \sum_{n=0}^{{N_\mathrm{t}}\!-\!2} \frac{{\left(g_k^\mathrm{th}\right)}^n}{n!} \!+\! e^{-g_k^\mathrm{th}} \frac{{\left(g_k^\mathrm{th}\right)}^{{N_\mathrm{t}}\!-\!2}}{\left({N_\mathrm{t}}\!-\!2\right)!}. \nonumber
\end{align}
We can prove that $F_{N_\mathrm{t}}\!\left(g_k^\mathrm{th}\right)$ is an upper bound of $B_{N_\mathrm{t}}\!\left( g_k^\mathrm{th} \right)$. The proof is omitted due to lack of space. Since the upper bound is in closed-form, we apply it as the QoS constraint imposed on the packet dropping probability.


While the optimal combination of $\varepsilon_k^c$, $\varepsilon_k^q$ and $\varepsilon_k^h$ will improve EE, the optimization leads to an intractable resource allocation problem but with marginal gain  \cite{She2016CrossLayer}. In the sequel, we consider $\varepsilon_k^c \!=\! \varepsilon_k^q \!=\! \varepsilon_k^h \!=\! \varepsilon_\mathrm{D}/3$. Then, the optimization problem can be formulated as
\begin{align}	\label{eq:problemAvePtot}
\mathop \mathrm{minimize} \limits_{\begin{subarray}{c}P_k^\mathrm{th}, W_k, N_\mathrm{t} \end{subarray}} &\quad \frac{1}{\rho} \sum_{k=1}^K {\mathbb{E}\left\{P_k^\mathrm{t}\right\}} + P^\mathrm{ca} N_\mathrm{t} + P_0^\mathrm{c} \\
\text{s.t.} &\quad \gamma_k \geq \exp\left\{\frac{l_k(\varepsilon_{\rm D}/3)}{W_k}+\frac{v_k(\varepsilon_{\rm D}/3)}{\sqrt{W_k}}\right\}-1, \label{eq:SNRreq}\tag{\theequation a} \\
& \quad F_{N_\mathrm{t}}\!\left(g_k^\mathrm{th}\right) \leq \varepsilon_\mathrm{D}/3, \label{eq:Drop}\tag{\theequation b} \\
& \quad {g_k^\mathrm{th}}=\frac {N_0 W_k \gamma_k} {\alpha_k {P_k^\mathrm{th}}}, \label{eq:Pth}\tag{\theequation c}\\
& \quad \sum_{k=1}^{K}{P^\mathrm{th}_k} \leq P^\mathrm{t}_{\max}, \quad  \sum_{k=1}^{K}{W_k} \leq W_\mathrm{max}, \label{eq:Wmax}\tag{\theequation d} \\
& \quad P_k^\mathrm{th} > 0, \; W_k >0, \; N_\mathrm{t} \geq 2, \; k=1,2,...,K,\nonumber
\end{align}
where \eqref{eq:SNRreq} is the constraint on $\varepsilon_k^c$ and $(D^q_{\max},\varepsilon_k^q)$ obtained by substituting $V_k\!=\!1$ and $\varepsilon^c_k \!=\! \varepsilon^q_k \!=\! \varepsilon_\mathrm{D}/3$ into \eqref{eq:SNR}, \eqref{eq:Drop} is the constraint on $\varepsilon^h_k$ since $F_{N_\mathrm{t}}\!\left( g_k^\mathrm{th} \right)$ is an upper bound of $B_{N_\mathrm{t}}\!\left( g_k^\mathrm{th} \right)$, \eqref{eq:Pth} reflects the relation of $P_k^{\rm th}$ and $g_k^{\rm th}$ in \eqref{eq:Pt1}, and $W_\mathrm{max}$ is the total bandwidth.

 After substituting $\gamma_k$ in \eqref{eq:Pth} into \eqref{eq:SNRreq}, we can see that the feasible region of the problem is non-convex. This is because the achievable rate with finite blocklength in \eqref{eq:sk} is non-convex. As a consequence, it is very challenging to find the global optimal solution.

\subsection{Optimal Resource Allocation Policy}
In this subsection, we propose a method to find the global optimal solution of problem \eqref{eq:problemAvePtot}. It is not hard to prove that $\mathbb{E}\left\{P_\mathrm{tot}\right\}$ increases with $P_k^\mathrm{th}$ given $W_k$ and $N_\mathrm{t}$. Moreover, according to the definition of $g_k^\mathrm{th}$, $P_k^\mathrm{th}$ decreases with $g_k^\mathrm{th}$. Hence, minimizing $\mathbb{E}\left\{P_\mathrm{tot}\right\}$ is equivalent to maximizing $g_k^\mathrm{th}$. Furthermore, we can prove that $F_{N_\mathrm{t}}\!\left(g_k^\mathrm{th}\right)$ increases with $g_k^\mathrm{th}$ (the proof is omitted due to the lack of space). Therefore, the maximal ${g_k^\mathrm{th}}$ can be obtained from $F_{N_\mathrm{t}}\!({g_k^\mathrm{th}}) \!=\! \varepsilon_{\rm{D}}/3$, and the minimal $P_k^\mathrm{th}$ with given $W_k$ and $N_\mathrm{t}$ can be expressed as
\begin{align}\label{eq:optP}
{P_k^\mathrm{th}}\!=\!\frac {N_0 W_k \gamma_k} {\alpha {g_k^\mathrm{th}}},\;\text{if}\; \sum_{k=1}^{K}{{P^\mathrm{th}_k}} \!\leq\! P^\mathrm{t}_{\max},
\end{align}
where $\gamma_k$ is the minimal value that satisfies constraint \eqref{eq:SNRreq}.
If $\sum_{k=1}^{K}{{P^\mathrm{th}_k}} > P^\mathrm{t}_{\max}$, we need to increase $N_{\rm t}$ until the power constraint satisfies. By substituting the optimal $W_k$ and $N_\mathrm{t}$ into \eqref{eq:optP}, optimal $P_k^\mathrm{th}$ can be obtained, and hence we optimize $W_k$ and $N_{\rm t}$ in the sequel.

Substituting $F_{N_\mathrm{t}}\!({g_k^\mathrm{th}}) \!=\! \varepsilon_{\rm{D}}/3$ into \eqref{eq:AvePt2}, we have
\begin{align}   \label{eq:AvePtApprox}
	\mathbb{E}\left\{P_k^\mathrm{t}\right\}=\frac{N_0 W_k \gamma_k \left( 1-\varepsilon_\mathrm{D}/3 \right)}{\alpha_k \left({N_\mathrm{t}}-1\right)}.
\end{align}
Substituting \eqref{eq:AvePtApprox} into \eqref{eq:AvePtot}, we can obtain that
\begin{align}\label{eq:avePtot}
\mathbb{E}\left\{P_\mathrm{tot}\right\} = \frac{N_0 \sum_{k=1}^K \!{\frac{W_k \gamma_k}{\alpha_k}} \left( 1\!-\! \varepsilon_\mathrm{D}/3 \right)}{\rho \left(N_\mathrm{t}-1\right)} + P^\mathrm{ca} N_\mathrm{t} + P_0^\mathrm{c}.
\end{align}
The optimal value of $W_k$ that minimizes $\mathbb{E}\left\{P_\mathrm{tot}\right\}$ is the same as that minimizes $\sum_{k=1}^K {\frac{W_k \gamma_k}{\alpha_k}}$, which does not depend on $N_{\rm t}$. However, the optimal value of $N_{\rm t}$ that minimizes $\mathbb{E}\left\{P_\mathrm{tot}\right\}$ depends on $W_k$. Therefore, we first find the optimal bandwidth allocation $W_k^*$, and then optimize $N_{\rm t}$ given $W_k^*$ in the sequel.

Since \eqref{eq:avePtot} increases with $\gamma_k$, the minimal $\mathbb{E}\left\{P_\mathrm{tot}\right\}$ is obtained when the equality of \eqref{eq:SNRreq} holds. Denote
\begin{align}	\label{eq:y}
	y_k\left(W_k\right)= W_k\gamma_k= W_k\left(e^{\frac{l_k (\varepsilon_{\rm D}/3)}{W_k}+\frac{v_k (\varepsilon_{\rm D}/3)}{\sqrt{W_k}}}-1\right).
\end{align}
Then, the optimal bandwidth allocation that minimizes $\mathbb{E}\left\{P_\mathrm{tot}\right\}$ under the QoS constraints can be found from the following problem,
\begin{align}	\label{eq:problemY}
	\mathop \mathrm{minimize} \limits_{{W_k},k=1,2,...,K} &\quad \sum_{k=1}^K {\frac{y_k\left(W_k\right)}{\alpha_k}}, \\
			\mathrm{s.t.}&\quad \sum_{k=1}^K {W_k} \leq W_\mathrm{max}\;\text{and}\;W_k > 0. \nonumber
\end{align}
$y_k\left(W_k\right)$ is non-convex in $W_k$. Essentially, this results from the non-convex achievable rate in \eqref{eq:sk}. Fortunately, we can exploit the following properties to find the global optimal solution.

\begin{pro} \label{P:monotone}
\emph{$y_k\left(W_k\right)$ first strictly decreases and then strictly increases with $W_k$. (See proof in Appendix \ref{App:yProp}.)}
\end{pro}
Property\ref{P:monotone} indicates that a unique solution $W_k^\mathrm{th}$ that minimizes $y_k\left(W_k\right)$ can be obtained via binary search algorithm.
\begin{pro} \label{P:convexity}
\emph{$y_k\left(W_k\right)$ is strictly convex in $W_k$ when $W_k \in (0,W_k^\mathrm{th})$. (See proof in Appendix \ref{App:yProp}.)}
\end{pro}

With these properties, we consider two complementary cases to find the global optimal solution of problem \eqref{eq:problemY}.

%

\subsubsection{Case 1}
In the case where the bandwidth is sufficiently large such that $\sum_{k=1}^K {W_k^\mathrm{th}} \! \leq \! W_\mathrm{max}$, $\{W_1^\mathrm{th},...,W_K^\mathrm{th}\}$ is a feasible solution  of problem \eqref{eq:problemY}. According to Property \ref{P:monotone}, $y_k\left(W_k\right)$ is minimized at $W_k^\mathrm{th}$, and hence the objective function \eqref{eq:problemY} is minimized at $\{W_1^\mathrm{th},...,W_K^\mathrm{th}\}$. Therefore, $\{W_1^\mathrm{th},...,W_K^\mathrm{th}\}$ is the global optimal solution, i.e., $W_k^* =W_k^\mathrm{th}$.

\subsubsection{Case 2}
In the case where $\sum_{k=1}^K {W_k^\mathrm{th}} \!>\! W_\mathrm{max}$, $\{W_1^\mathrm{th},\dotsc,W_K^\mathrm{th}\}$ is not a feasible solution of problem \eqref{eq:problemY}. However, with Property \ref{P:monotone}, it is not hard to see that the global optimal bandwidth allocation satisfies $W_k^* \!\leq\! W_k^\mathrm{th}$. Then, problem \eqref{eq:problemY} is equivalent to the following problem,
\begin{align}	\label{eq:problemYConvex}
	\mathop \mathrm{minimize} \limits_{W_k,k=1,2,...,K} &\quad \sum_{k=1}^K {\frac{y_k\left(W_k\right)}{\alpha_k}} \\
			\mathrm{s.t.} 					   &\quad \sum_{k=1}^K {W_k} = W_\mathrm{max}, \; 0 < {W_k} \leq W^\mathrm{th}_{k}, \nonumber
\end{align}
which is convex according to Property \ref{P:convexity}. Then, the global optimal bandwidth allocation $W^*_k$ can be obtained with interior point method.

By substituting $W^*_k$ into  \eqref{eq:avePtot}, the global optimal number of active transmit antennas that minimizes $\mathbb{E}\left\{P_\mathrm{tot}\right\}$ under the QoS constraints can be derived as
\begin{align}
	N_\mathrm{t}^*=\left\lceil \frac{1}{2} \left( 1 + \sqrt{ 1 + \frac{4 N_0 \left( 1-\varepsilon_\mathrm{D}/3 \right) \sum_{k=1}^K {\frac{y_k\left(W^*_k\right)}{\alpha_k}}}{\rho P^\mathrm{ca}} } \right) \right\rceil, \nonumber
\end{align}
where $\lceil x \rceil$ denotes the minimal integer no less than $x$.

\section{Simulation and Numerical Results}
In this section, we validate the approximation and upper bound introduced to ensure the QoS, illustrate properties \ref{P:monotone} and \ref{P:monotone} and evaluate the EE achieved by the proposed policy.

We set $\varepsilon_k^c\!=\!\varepsilon_k^q\!=\!\varepsilon_k^h\!=\!{\varepsilon_\mathrm{D}}/{3} \!=\! \!10^{-7}$. The users are uniformly distributed in the range of $50$~m $\sim$ $250$~m away from the BS. There are $20$ nodes lie in the concerned area of each user, which may be sensors or other users around the target user. The average packet arrival rate from each node is $10$ packets per second, which is the typical value in vehicle networks as well as some other use  cases\cite{Philipp2017Latency}. Other parameters are listed in Table \ref{tab:SimParam}.

\begin{table}[htbp]
	\vspace{-0.4cm}
	\small
	\renewcommand{\arraystretch}{1.3}
	\caption{Simulation Parameters}	\label{tab:SimParam}
	\begin{center}\vspace{-0.2cm}
	\begin{tabular}{|p{5.3cm}|p{2.4cm}|}
		\hline
        E2E delay requirement $D_{\max}$ & $1$~ms \\ \hline
		Duration of each frame $T_\mathrm{f}$ & $0.1$~ms \\ \hline
		Duration of DL transmission  $\phi$ & $0.05$~ms \\ \hline
        Backhaul latency & $0.1$~ms \cite{Tony2015Delay} \\ \hline
        Queueing delay requirement $D^q_{\max}$ & $0.8$~ms \\ \hline
		Single-sided noise spectral density $N_0$ & $-173$~dBm/Hz \\ \hline
		Available bandwidth $W_\mathrm{max}$ & $20$~MHz \\ \hline
		Packet size $u$ & $160$~bits \cite{3GPP2016Scenarios} \\ \hline
		Path loss model $10\lg(\alpha_k)$ & $35.3+37.6 \lg(d_k)$ \\ \hline
        Maximal transmit power of BS $P^\mathrm{t}_\mathrm{max}$ & $40$~dBm \\ \hline
        Circuit power consumed per antenna $P^\mathrm{ca}$ & $50$~mW \cite{Bjorn2015A} \\ \hline
        Fixed circuit power $P_0^\mathrm{c}$ & $50$~mW \cite{Bjorn2015A} \\ \hline
        power amplifier efficiency $\rho$ & $0.5$ \cite{Bjorn2015A}\\ \hline
	\end{tabular}
	\end{center}
	\vspace{-0.4cm}
\end{table}

\begin{table}[htbp]
	\vspace{-0.2cm}
	\small
	\renewcommand{\arraystretch}{1.3}
	\caption{Validation of the Approximation and Bound} \label{tab:Validation}
	\begin{center}\vspace{-0.2cm}
    \begin{tabular}{|@{ }c@{ }|@{ }c@{ }|@{ }c@{ }|@{ }c@{ }|@{ }c@{ }|} \hline
        Required $\varepsilon_k^h$ & $10^{-8}$ & $10^{-7}$ & $10^{-6}$ & $10^{-5}$ \\ \hline
        Achieved $\varepsilon_k^h$ & $1.9\!\times\!10^{-9}$ & $9.2\!\times\!10^{-9}$ & $1.8\!\times\!10^{-7}$ & $3.9\!\times\!10^{-6}$ \\ \hline
	\end{tabular}
	\end{center}
	\vspace{-0.4cm}
\end{table}

The impact of using the approximated $\varepsilon^h$ in \eqref{eq:UB} and the upper bound of $B_{N_\mathrm{t}}\!\left(g_k^\mathrm{th}\right)$ (i.e., $F_{N_\mathrm{t}}\!\left(g_k^\mathrm{th}\right)$) in the optimization is shown in Table \ref{tab:Validation}, where $K=1$, $M=20$ and the user-BS distance is $250$~m. The achieved $\varepsilon_k^h$ is obtained by computing the average number of discarded packets from \eqref{dk} with the optimal policy over $10^{11}$ Rayleigh fading channel realizations and then substituting it into \eqref{eq:UB}. We can see that the achieved $\varepsilon^h$ are lower than the required $\varepsilon^h$, which means that the QoS requirement can be guaranteed with the approximation and upper bound used in constraint \eqref{eq:Drop}. Similar results have been obtained for $K\!\geq\!2$ that give rise to the same conclusion, which are omitted for conciseness. By substituting $F_{N_\mathrm{t}}\!({g_k^\mathrm{th}}) \!=\! \varepsilon_k^h$ into \eqref{eq:AvePt2}, we can see that the average transmit power is insensitive to proactive packet dropping probability, i.e., the approximation and the bound have minor impact on EE.

\begin{table}[htbp]
	\vspace{-0.2cm}
	\small
	\renewcommand{\arraystretch}{1.3}
	\caption{Values of $W^\mathrm{th}$ that Minimize $y_k\left(W_k\right)$} \label{tab:Wth}
	\begin{center}\vspace{-0.2cm}
    \begin{tabular}{|@{\quad}c@{\quad}|@{\quad}c@{\quad}|@{\quad}c@{\quad}|@{\quad}c@{\quad}|@{\quad}c@{\quad}|} \hline
        $\varepsilon_k^c$ & $10^{-8}$ & $10^{-7}$ & $10^{-6}$ & $10^{-5}$ \\ \hline
        $W^\mathrm{th}$~(MHz) & $7.35$ & $7.42$ & $7.53$ & $7.70$ \\ \hline
	\end{tabular}
	\end{center}
	\vspace{-0.4cm}
\end{table}

Values of $W^\mathrm{th}$ with different transmission error probability requirements are shown in Table \ref{tab:Wth}, where $E^\mathrm{B} \!=\! 1$. It can be seen that $W^\mathrm{th}$ is much larger than the coherence bandwidth (e.g., $0.5$~MHz). Furthermore, we can be prove that $W^\mathrm{th}$ increases with $E^\mathrm{B}$ and $u$, and does not change with the channel gain. The proof is omitted due to lack of space. Thus, \eqref{eq:y} strictly decreases with $W_k$ and is convex in $W_k$ if $W_k$ is smaller than coherence bandwidth.

\begin{figure}[htbp]
	\vspace{-0.2cm}
	\centering
	\begin{minipage}[t]{0.42\textwidth}
	\includegraphics[width=1\textwidth]{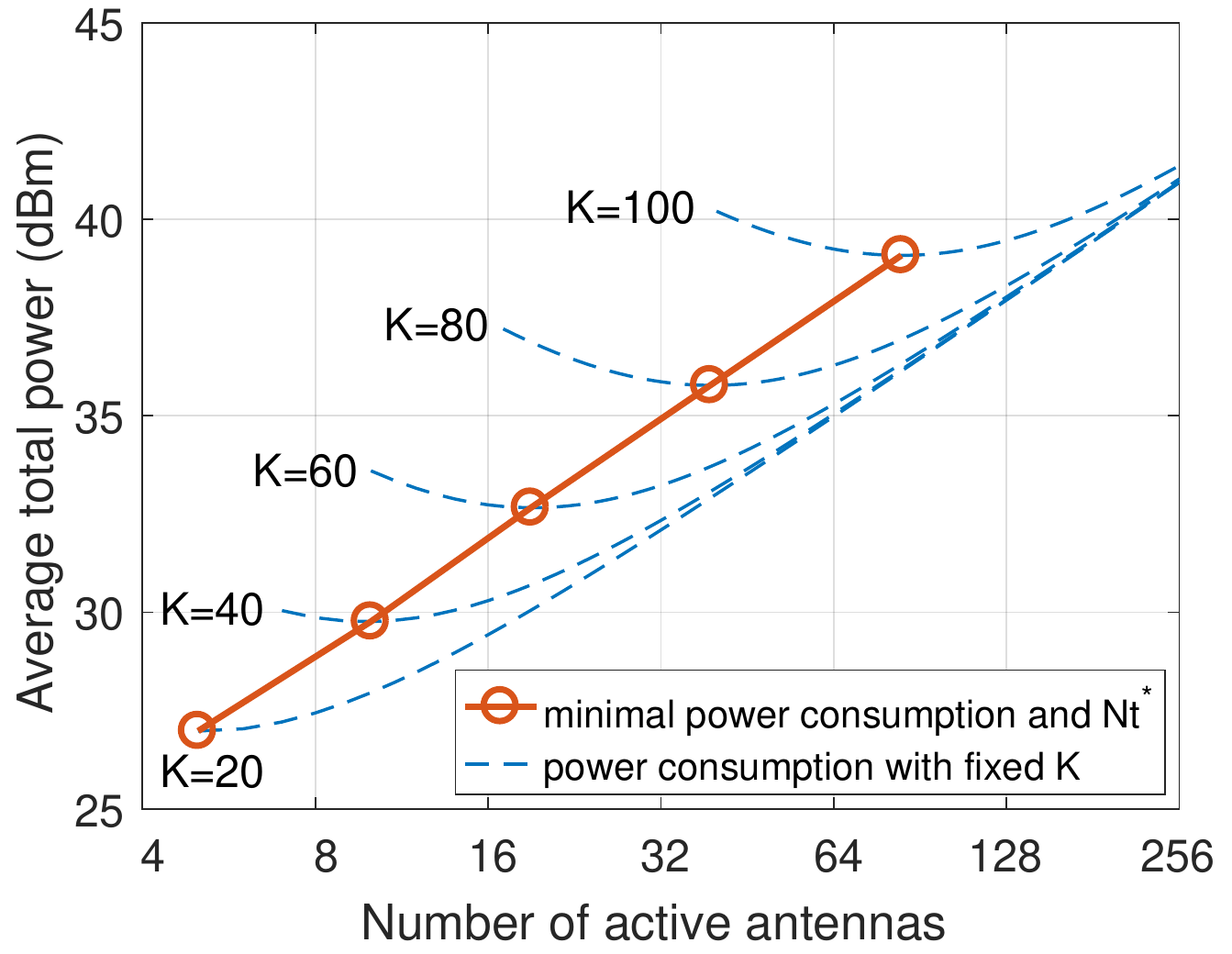}
	\end{minipage}
	\vspace{-0.2cm}
	\caption{Average total power consumption v.s. active number of antennas.}	\label{fig:Ptot}
	\vspace{-0.2cm}
\end{figure}
The average total power consumption obtained with different numbers of active transmit antennas is shown in Fig. \ref{fig:Ptot}. The dash curves are obtained with different number of users,  which reflect the traffic load. It shows that the average total power consumption first decreases and then increases with $N_\mathrm{t}$ for given user density. This is because there is a tradeoff between transmit power and circuit power. Transmit power is high when $N_\mathrm{t}$ is small, and circuit power is high when $N_\mathrm{t}$ is large. The solid curve shows the relation between the minimal average total power consumption and the corresponding optimal number of active transmit antennas. It shows that both the minimum average total power and $N_\mathrm{t}^*$ increase  with $K$.

\begin{figure}[htbp]
	\vspace{-0.2cm}
	\centering
	\begin{minipage}[t]{0.45\textwidth}
	\includegraphics[width=1\textwidth]{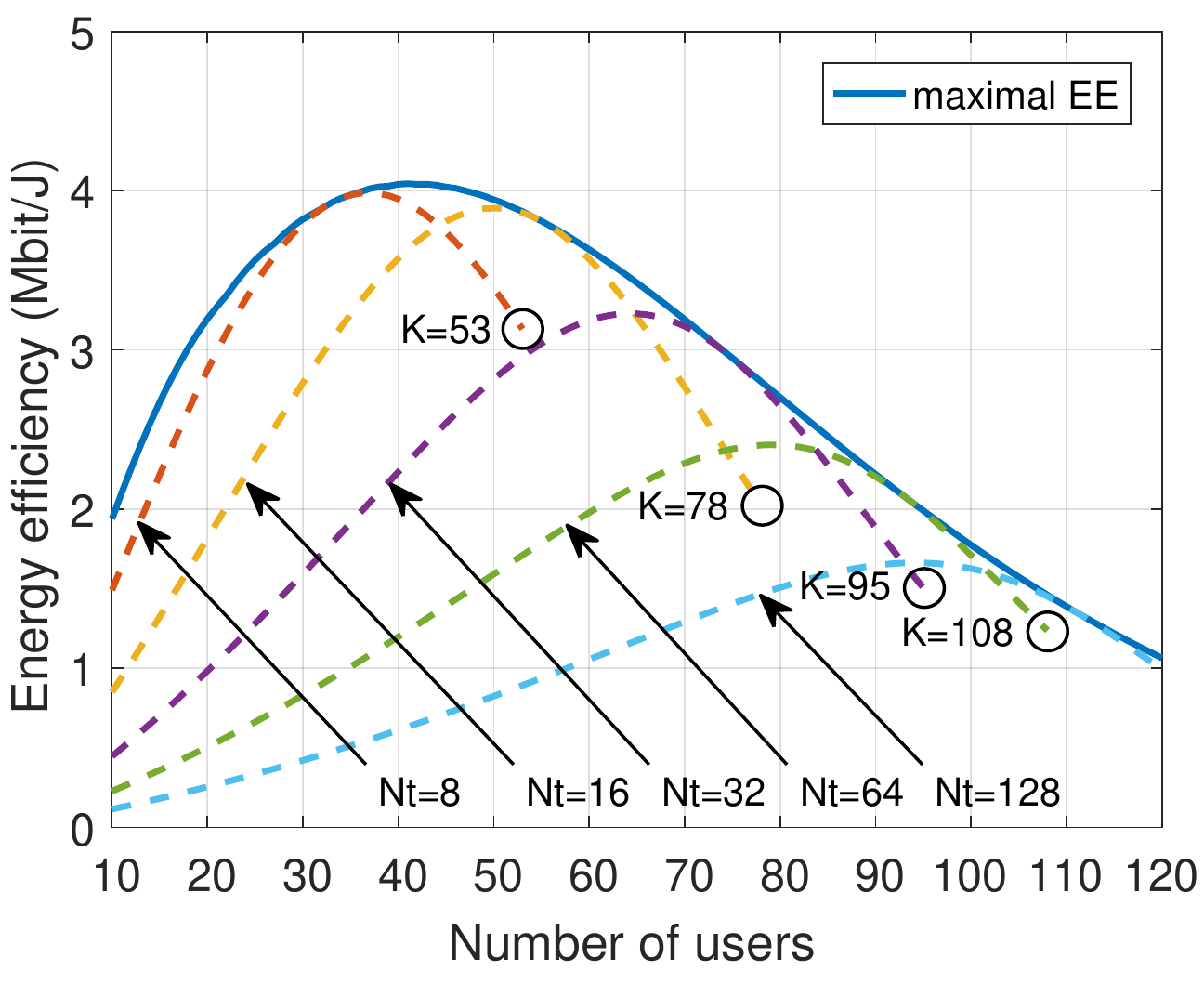}
	\end{minipage}
	\vspace{-0.2cm}
	\caption{Maximal EE v.s. the number of users.}	\label{fig:EE}
	\vspace{-0.2cm}
\end{figure}
The maximized EE versus number of users is demonstrated in Fig. \ref{fig:EE}. The dash curves are the maximal EE achieved by optimizing transmit power and bandwidth allocation with given number of antennas (which can be regarded as a modified policy of \cite{She2016EEtactile} by considering a more accurate approximation of achievable rate), where the circle in each curve indicates the maximal number of users that can be supported with the given resources. The solid curve is the maximal EE achieved by the proposed optimal resource allocation, where the number of antennas is jointly optimized with transmit power and bandwidth. As is shown, the EE increases with $K$ under light traffic load, but decreases with $K$ under heavy traffic load. The gaps between the solid curve and the dash curves indicate that adjusting the number of active antennas according to the number of users is critical to maximizing the EE.

\section{Conclusion}
In this paper, we studied how to optimize energy efficient resource allocation for URLLC. We formulated a problem to optimize transmit power, bandwidth, and number of active antennas that maximizes the EE under the ultra-low E2E latency and ultra-high overall reliability requirements. Although the optimization problem is non-convex owing to the non-convex achievable rate with finite blocklength channel codes, the global optimal solution was obtained. Simulation and numerical results validated our analysis and showed that by jointly optimizing the number of active antennas together with the transmit power and bandwidth, the EE of the system can be improved significantly.

\appendices

\section{Proof of the properties of \eqref{eq:y}}
\label{App:yProp}
\renewcommand{\theequation}{A.\arabic{equation}}
\setcounter{equation}{0}
\begin{proof}
The index $k$ is omitted in this appendix for notational simplicity. The first order and second order derivatives of $y(W)$ are respectively
\begin{align}
	y'(W) =&\left(\!1\!-\!\frac{l}{W}\!-\!\frac{v}{2\sqrt{W}}\right) e^{\!\frac{l}{W}\!+\!\frac{v}{\sqrt{W}}} \!-\! 1, \\
	y''(W)=&\left(\!-vW^{\sfrac{3}{2}}\!+\!v^2 W\!+\!4lv\sqrt{W}\!+\!4l^2\!\right) \frac{e^{\!\frac{l}{W}\!+\!\frac{v}{\sqrt{W}}}}{4W^3}. \label{eq:ydd}
\end{align}
Denote $x(W) = -vW^{\sfrac{3}{2}}+v^2 W+4lv\sqrt{W}+4l^2$. Notice that $y''(W)\!=\!0$ if and only if $x(W)\!=\!0$. We can also obtain that  $x'(W) \!>\! 0$, $W \!\in\! [0,W^\mathrm{(1)})$ and $x'(W) \!<\! 0$, $W \!\in\! (W^\mathrm{(1)},+\infty)$, where $W^\mathrm{(1)}$ is the solution of $x'(W^\mathrm{(1)}) \!=\! 0$. Since $x(0) \!=\! 4l^2 \!>\! 0$ and $x(W)$ increases with $W$ when $W \!\in\! [0,W^\mathrm{(1)})$, $x(W) \!>\! 0, \forall \; W \!\in\! [0,W^\mathrm{(1)})$. Moreover, $x(W)$ decreases with $W$ when $W \!\in\! (W^\mathrm{(1)},+\infty)$, and $x(W) \!\to\! -\! \infty$ when $W \!\to\! \infty$. Hence, we can find $W^\mathrm{(0)}$ in $(W^\mathrm{(1)},+\infty)$, which is the solution of $x(W^\mathrm{(0)})\!=\!0$. Then, we have
\begin{align}\label{eq:secondorder}
	y''(W) \begin{cases}
		>0 &\quad \text{if } 0<W<W^\mathrm{(0)}, \\
		=0 &\quad \text{if } W=W^\mathrm{(0)}, \\
		<0 &\quad \text{if } W>W^\mathrm{(0)}.
	\end{cases}
\end{align}

By analyzing \eqref{eq:ydd}, we can find $W^\mathrm{th} \!\in\! (0,W^\mathrm{(0)})$ that satisfies $y'(W^\mathrm{th}) \!=\! 0$ (the details are omitted due to the lack of space). Then, from \eqref{eq:secondorder} and $\lim_{W \to \infty}y'(W)\!=\!0$ we have
\begin{align}
	y'(W) \begin{cases}
		<0 &\quad \text{if } 0<W<W^\mathrm{th}, \\
		=0 &\quad \text{if } W=W^\mathrm{th}, \\
		>0 &\quad \text{if } W^\mathrm{th} <W .
	\end{cases}
\end{align}
Therefore, there exists a unique point $W^\mathrm{th} \!\in\! ({0,+\infty})$ that minimizes $y(W)$. Since $y'(W)\!<\!0$ and $y''(W)\!>\!0$ when  $W \in ({0,W^\mathrm{th}})$, $y(W)$ is a strict decreasing and convex function of $W$ when $W \!\in\! ({0,W^\mathrm{th}})$.
\end{proof}

\bibliographystyle{IEEEtran}
\bibliography{ref}

\begin{thebibliography}{10}
\providecommand{\url}[1]{#1}
\csname url@samestyle\endcsname
\providecommand{\newblock}{\relax}
\providecommand{\bibinfo}[2]{#2}
\providecommand{\BIBentrySTDinterwordspacing}{\spaceskip=0pt\relax}
\providecommand{\BIBentryALTinterwordstretchfactor}{4}
\providecommand{\BIBentryALTinterwordspacing}{\spaceskip=\fontdimen2\font plus
\BIBentryALTinterwordstretchfactor\fontdimen3\font minus
  \fontdimen4\font\relax}
\providecommand{\BIBforeignlanguage}[2]{{%
\expandafter\ifx\csname l@#1\endcsname\relax
\typeout{** WARNING: IEEEtran.bst: No hyphenation pattern has been}%
\typeout{** loaded for the language `#1'. Using the pattern for}%
\typeout{** the default language instead.}%
\else
\language=\csname l@#1\endcsname
\fi
#2}}
\providecommand{\BIBdecl}{\relax}
\BIBdecl

\bibitem{Factory2015Yilmaz}
{O. N. C. Yilmaz, Y.-P. E. Wang, N. A. Johansson, \emph{et al.}}, ``Analysis of
  ultra-reliable and low-latency 5{G} communication for a factory automation
  use case,'' in \emph{IEEE ICC Workshops}, 2015.

\bibitem{Meryem2016Tactile}
{M. Simsek, A. Aijaz, M. Dohler, \emph{et al.}}, ``5{G}-enabled tactile
  internet,'' \emph{IEEE J. Select. Areas Commun.}, vol.~34, no.~3, pp.
  460--473, Mar. 2016.

\bibitem{Petteri2015A}
{P. Kela, J. Turkka, \emph{et al.}}, ``A novel radio frame structure for 5{G}
  dense outdoor radio access networks,'' in \emph{Proc. IEEE VTC Spring}, 2015.

\bibitem{Adnan2016Towards}
A.~Aijaz, ``Towards 5{G}-enabled tactile internet: Radio resource allocation
  for haptic communications,'' in \emph{Proc. IEEE WCNC}, 2016.

\bibitem{She2016CrossLayer}
C.~She, C.~Yang, and T.~Q. Quek, ``Cross-layer optimization for ultra-reliable
  and low-latency radio access networks,'' \emph{IEEE Trans. Wireless Commun.,
  revised}, \url{https://arxiv.org/pdf/1703.09575.pdf}.

\bibitem{Yury2014Quasi}
{W. Yang, G. Durisi, T. Koch, \emph{et al.} }, ``Quasi-static multiple-antenna
  fading channels at finite blocklength,'' \emph{IEEE Trans. Inf. Theory},
  vol.~60, no.~7, pp. 4232--4264, Jul. 2014.

\bibitem{Beatriz2015Reliable}
{G. Pocovi, B. Soret, M. Lauridsen, \emph{et al.}}, ``Signal quality outage
  analysis for ultra-reliable communications in cellular networks,'' in
  \emph{IEEE Globecom Workshops}, 2015.

\bibitem{YangGR2015}
G.~Wu, C.~Yang, S.~Li, and G.~Li, ``Recent advance in energy-efficient networks
  and its application in {5G} systems,'' \emph{IEEE Wireless Commun. Mag.},
  vol.~22, no.~2, pp. 145--151, Apr. 2015.

\bibitem{Wenjuan2016TWC}
W.~Yu, L.~Musavian, and Q.~Ni, ``Tradeoff analysis and joint optimization of
  link-layer energy efficiency and effective capacity toward green
  communications,'' \emph{IEEE Trans. on Wireless Commun.}, vol.~15, no.~5, pp.
  3339--3353, May 2016.

\bibitem{She2016EEtactile}
C.~She and C.~Yang, ``Energy efficient design for tactile internet,'' in
  \emph{Proc. IEEE/CIC ICCC}, 2016.

\bibitem{EB}
C.~Chang and J.~A. Thomas, ``Effective bandwidth in high-speed digital
  networks,'' \emph{IEEE J. Sel. Areas Commun.}, vol.~13, no.~6, pp.
  1091--1100, Aug. 1995.

\bibitem{Tony2015Delay}
{G. Zhang, T. Q. S. Quek, A. Huang, \emph{et. al.}}, ``Delay modeling for
  heterogeneous backhaul technologies,'' in \emph{Proc. IEEE VTC Fall}, 2015.

\bibitem{Philipp2017Latency}
{P. Schulz, M. Matth\'{e}, H. Klessig, \emph{et al.}}, ``Latency critical
  {I}o{T} applications in 5{G}: Perspective on the design of radio interface
  and network architecture,'' \emph{IEEE Commun. Mag.}, vol.~55, no.~2, pp.
  70--78, Feb. 2017.

\bibitem{3GPP2016Scenarios}
3GPP, \emph{Study on Scenarios and Requirements for Next Generation Access
  Technologies}.\hskip 1em plus 0.5em minus 0.4em\relax Technical Specification
  Group Radio Access Network, Technical Report 38.913, Release 14, Oct. 2016.

\bibitem{Bjorn2015A}
B.~Debaillie, C.~Desset, and F.~Louagie, ``A flexible and future-proof power
  model for cellular base stations,'' in \emph{Proc. IEEE VTC Spring}, 2015.

\end{thebibliography}

\end{document}